\documentclass[reprint,aps,pra,showpacs,
amsmath,amssymb]{revtex4-1}
\usepackage{mathtools}
\usepackage{graphicx}
\usepackage{booktabs,multirow}

\begin{document}
\newcommand{\pd}[2]{\frac{\partial #1}{\partial #2}} 
\newcommand{\ket}[1]{\left| #1 \right>} 
\newcommand{\bra}[1]{\left< #1 \right|} 
\newcommand{\bs}{\boldsymbol}
\newcommand{\sech}{\text{sech}}
\title{Theory of storage and retrieval of intense-broadband pulses at room 
temperature: Analytical and numerical solutions}

\author{Rodrigo Guti\'{e}rrez-Cuevas}
\email{rgutier2@ur.rochester.edu}
\affiliation{Center for Coherence and Quantum Optics and the Institute of 
Optics, University of Rochester, Rochester, New York 14627, USA}

\date{\today}

\begin{abstract}
We analyze the storage and retrieval of intense-broadband pulses with the 
added effects of Doppler broadening and detuning in a $\Lambda$ configuration. 
We compute analytical solutions via the inverse scattering technique and show 
how the signal field is transferred to a spin-wave in the atomic medium and 
later retrieved by the interaction of a control pulse. Due to the intensity of 
the pulses, the pulse area (as defined for self-induced transparency) plays a 
key role in the interaction, as it determines the location of the spin wave 
within the medium. Additionally, we compare our results to non-ideal 
conditions by considering pulses of finite length and the effect of 
spontaneous emission.
\end{abstract}


\maketitle

\section{Introduction}

The storage of light in atomic gases has its roots in photon echoes, where it 
was recognized that an intense $\pi$ pulse could reverse the free induction 
decay produced by an initial pulse with arbitrary area. This inversion leads 
to the emission of an echo of the original pulse 
\cite{hahn1950spin,kurnit1964observation,mossberg1982time}. The problem with 
the original work was that it was conceived for a two-level system for which 
the transverse relaxation time is quite short, thus significantly limiting 
storage time.  In addition, the shape of the echo is not the same as that of 
the stored pulse. 

\begin{figure}
\includegraphics[scale=1]{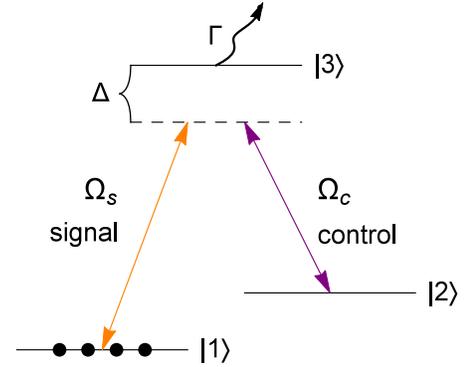}
\caption{\label{fig:lambda} (Color online) Three-level
 atom in $\Lambda$-configuration 
interacting with two fields in two-photon resonance via the common detuning 
$\Delta$, with spontaneous emission $\Gamma$ from the excited state.}
\end{figure}

All these issues have been addressed throughout the years and many techniques 
have been proposed, most of them based on the $\Lambda$ configuration (see 
Fig.~\ref{fig:lambda}). In this case, one can make the coherence between the 
two ground states to be long lived, thus increasing storage time. Most of the 
proposals are based on the phenomenon of electromagnetically induced 
transparency (EIT) 
\cite{harris1997electromagnetically,*boller1991observation}, where the  signal 
(or probe) pulse is slowed down by turning off the intense control field until 
it is stored in the ground states and subsequently restored by turning the 
control back on 
\cite{fleischhauer2000dark,matsko2001nonadiabatic,dey2003storage}.
In order to avoid any significant loss due to spontaneous decay from the 
higher level, schemes based on stimulated Raman transitions have been 
proposed. Due to the high detuning, the higher atomic level can be 
adiabatically eliminated, which leads to simpler equations 
\cite{nunn2007mapping,*reim2011single}.
Extensions to the photon echo technique have also been worked out, where the 
signal pulse is mapped into a spin wave of the ground states by means of an 
intense $\pi$-control pulse and then retrieved by another counter-propagating 
strong-control pulse \cite{moiseev2001complete}. 

All these techniques address the same problem: How can we store a given signal 
field into an atomic medium to then retrieve it? We call this an optical 
memory even if the retrieved pulse is not the same as the original one: it 
might have a different shape, time duration, or intensity. Of course, this 
might not matter if we are only looking at whether a pulse is there or not. 
Nonetheless, the ideal scenario is the one in which the retrieved pulse is 
identical to the one stored (i.e.,~has the same shape and parameters as the 
original signal pulse). In any case, when one talks about optimizing these 
types of memories, a figure of merit must be defined. In the series of papers 
\cite{gorshkov2007universal,*gorshkov2007photon,*gorshkov2007photon2, 
*gorshkov2008photon,gorshkov2007photon3} by Gorshkov \emph{et al.} they 
addressed this issue by defining the efficiency as the ratio of the retrieved 
pulse intensity to that of the original signal pulse. Hence they can find the 
optimal control field to store and retrieve a given signal field by maximizing 
this efficiency. This work as well as that by Nunn \emph{et al.} shows that, 
in general, the optimal control field has some temporal shape which differs 
from the usual on/off variation of the cw control used in EIT-based schemes.

Most works on pulse storage deal with a quantum (low intensity) signal pulse 
which allows a treatment up to first order in the signal field. There is also 
a recurring assumption of adiabaticity (this point has been addressed in 
\citep{matsko2001nonadiabatic,shakhmuratov2007instantaneous}) for the control 
pulse, and its spatial evolution is usually left out. These approximations are 
understandable as the full set is composed of eight nonlinear partial 
differential equations which resist analytic solutions except in special cases 
\cite{grobe1994formation,park1998matched,clader2007two,groves2013jaynes,
gutierrez2015multi}. 
Therefore numerical calculations are a recurring tool to study the full 
evolution 
\cite{matsko2001nonadiabatic,dey2003storage,gutierrez2015manipulation}.
In the present work, we move away from these assumptions and consider the full 
nonlinear equations for intense-short pulses (non-quantum and non-adiabatic 
fields). This is the regime of self-induced transparency (SIT), where, in a 
two-level system, a resonant-intense-$2\pi$ pulse propagates without 
absorption and preserving its shape \cite{mccall1967self,*mccall1969self}. An 
SIT pulse can be understood as a pulse that is constantly encoding and 
retrieving itself resulting in a reduced group velocity.

The storage and retrieval of intense broadband pulses in a $\Lambda$ system 
has been shown to work for cold atoms in 
\cite{groves2013jaynes,gutierrez2015manipulation}. This was further extended 
to add more control to the information stored and the storage of multiple 
pulses \cite{gutierrez2015multi}. Additionally, within the same framework, the 
possibility for two-channel memory and vector-pulse storage in a tripodal 
system has been discussed in \cite{gutierrez2016vector}. In these works, the 
storage of the intense signal pulse follows a similar scheme to the one 
proposed in \cite{grigoryan2009short} for storage (a weak control pulse and 
intense signal pulse) but they do not discuss any retrieval procedure. 

In the present work we treat the storage and retrieval of intense short pulses 
in the presence of Doppler broadening and detuning (this has been treated for 
the case of low-intensity light 
\cite{gorshkov2007photon3,sangouard2007analysis}). For this, we solve the 
evolution equations by means of the inverse scattering transform 
\cite{gardner1967method,*ablowitz1973nonlinear,*lamb1980elements,
chakravarty2014inverse} for which we present an operational summary based on 
the work  presented by Chakravarty \emph{et al.} \cite{chakravarty2014inverse} 
for the particular case of a $\Lambda$ system. The first and second order 
solutions studied show the storage and retrieval of a signal pulse while 
pointing towards the key parameters for this scheme, such as the pulse areas 
and their duration. We follow by a numerical exploration adding the effect of 
spontaneous emission and show that the scheme presented remains valid.

\section{Physical Model}

As was mentioned, we are looking at the storage of pulses in the 
characteristic regime of SIT: short (of the order of 1ns) and intense (areas 
around $2\pi$) pulses. Therefore, we can omit (for the time being) the effects 
of spontaneous decay. We assume the atomic gas to be sufficiently dilute so as 
to safely neglect any loss of  coherence due to collisions. Additionally, we 
consider that the fields are close to resonance thus validating the use of the 
rotating-wave approximation. Writing the fields in carrier-envelope form,
\begin{align}
\bs{E}(x,t)=&\bs{\mathcal E}_{s}(x,t)e^{-i(k_{s}x
-\omega_{s}t)}\nonumber\\
&+\bs{\mathcal E}_{c}(x,t)e^{-i(k_{c}x-\omega_{c}t)}+c.c,
\label{carrier}
\end{align} 
we have that the Hamiltonian for the $\Lambda$-configuration (see Fig.~
\ref{fig:lambda}) takes the form
\begin{equation}
\bs H=-\frac{\hbar}{2} \left(
\begin{array}{ccc} 
0&0&\Omega_{s}\\
0&0&\Omega_{c}\\
\Omega_{s}^*&\Omega_{c}^*&-2\Delta
\end{array}
\right).
\label{hrwa}
\end{equation}
Here, we defined the Rabi frequencies for the signal and control fields, $
\Omega_{s,c}(x,t)=2\bs{d}_{s,c}\cdot \bs{\mathcal E}_{s,c}(x, t )/\hbar$ 
($d_{s,c}$ denotes the component of the dipole moment for each transition), 
and the detuning $\Delta=\omega_{31}-\omega_{s}=\omega_{32}-\omega_{c}$ for 
the two-photon resonance case. 

We are interested in studying the joint evolution of the atomic system and the 
two fields. Hence, the equations that determine the interaction are the von 
Neumann equation for the atomic medium,
\begin{equation}
i\hbar \pd{\bs \rho}{t}=[\bs H,\bs \rho],
\label{neumann}
\end{equation}
and the wave equation in the slowly varying envelope approximation for the 
fields,
\begin{subequations}
\label{meqs}
\begin{align}
\left(\pd{ }{x}+\frac{1}{c}\pd{}{t}\right)\Omega_{s}&=-i\mu_{s} \langle
\rho_{13}\rangle,\\
\left(\pd{ }{x}+\frac{1}{c}\pd{}{t}\right)\Omega_{c}&=-i\mu_{c} \langle 
\rho_{23} \rangle.
\end{align}
\end{subequations}
Here we defined the atom-field coupling parameters $\mu_{s,c}=N\omega_{s,c}|
d_{s,c}|^2/\hbar \epsilon_0c$ and the brackets denote the Doppler averaging, 
\begin{align}
\langle \rho \rangle = \int \rho(\Delta) F(\Delta) d\Delta,
\end{align}
over the frequency distribution 
\begin{align}
F(\Delta)= \frac{T_2^\star}{\sqrt{2 \pi}}e^{-(\Delta-\bar\Delta)^2T_2^{\star 
2}/2}
\end{align}
where $T_2^\star$ is the Doppler lifetime and $\bar\Delta$ is the mean one-
photon detuning.
This set of partial differential equations is referred to as the coupled 
Maxwell-Bloch (CMB) equations.

By further assuming equal atom-field parameters the CMB equations become 
integrable and can be written in the following form:
\begin{subequations}
\label{eq:mb}
\begin{equation}
 \pd{\bs \rho}{T}=[\bs \Omega -i\frac{\Delta}{2} \bs J,\bs \rho] 
\end{equation}
and
\begin{equation}
\pd{\bs \Omega}{Z}=\frac{\mu}{4}[\langle\bs \rho \rangle,\bs J].
\end{equation}
\end{subequations}
We introduced the traveling-wave coordinates $T=t-x/c$ ($c$ being the speed of 
light in vacuum) and $Z=x$, and defined the matrices
\begin{align}
\bs J=\left(
\begin{array}{ccc} 
-1&0&0\\
0&-1&0\\
0&0&1
\end{array}
\right), \quad 
\bs \Omega=\frac{i}{2} \left(
\begin{array}{ccc} 
0&0&\Omega_{s}\\
0&0&\Omega_{c}\\
\Omega_{s}^*&\Omega_{c}^*&0
\end{array}
\right).
\end{align}
The integrability of the system allows the use of standard methods to find 
analytic solutions. Some of these methods are  inverse scattering 
\cite{gardner1967method,*ablowitz1973nonlinear,*lamb1980elements,
chakravarty2014inverse},  the 
B\"acklund transformation 
\cite{lamb1971analytical,*miura1976backlund,*park1998field} and 
the Darboux transformation \cite{gu2006darboux,*cieslinski2009algebraic}.

\section{Inverse Scattering}

The method of inverse scattering allows the incorporation of Doppler 
broadening in a natural way for multi-pulse solutions, so we will use it to 
compute analytical solutions to the CMB equations instead of the B\"acklund or 
Darboux  transformations used in previous works 
\cite{park1998matched,clader2007two,clader2008two,groves2013jaynes,
gutierrez2015multi}. In what follows, we give a pragmatic presentation of the 
method without going into a careful derivation of the results. We refer the 
interested reader to the work by Chakravarty \emph{et al.~}
\cite{chakravarty2014inverse} for a complete presentation of the method for 
the case of a $\Lambda$ system. 

Since the CMB equations [Eqs.~\eqref{eq:mb}] are integrable, they can be 
expressed as the integrability condition of the linear system 
\begin{subequations}
\begin{align}
\pd{\bs\varphi}{T}&=\left(-\frac{i\lambda}{2} \bs J+\bs \Omega\right)\bs
\varphi \label{eq:scat1} \\
\pd{\bs\varphi}{Z}&=\frac{i\mu}{2}\left\langle\frac{\bs \rho}{\lambda-\Delta} 
\right\rangle \bs\varphi \label{eq:scat2}.
\end{align}
\end{subequations}
That is, we recover Eqs.\eqref{eq:mb} by collecting terms with the same $
\lambda$-dependence from the equation $\partial_Z\partial_T\bs\varphi=
\partial_T\partial_Z\bs\varphi$. The parameter $\lambda$ is referred to as the 
spectral parameter.

We assume the boundary conditions $\bs \Omega \rightarrow 0$ as $T \rightarrow 
\pm \infty$ for all $Z$, meaning the fields are composed of well-defined 
pulses, and fix the value of the density matrix at $T \rightarrow - \infty$ 
which we denote by $\bs \rho^{(0)}$.
This, in turn, will define the final state of the atomic system $\bs \rho^{f}=
\bs \rho(T \rightarrow \infty)$.
 
By integrating Eq.~\eqref{eq:scat1} we can define two solutions 
\begin{subequations}
\label{eq:phisi}
\begin{align}
\boldsymbol\Phi(T,\lambda)&=e^{-i\frac{\lambda}{2}\bs J T}+\int_{-\infty}^T 
e^{-i\frac{\lambda}{2}\bs J(T-T')}\boldsymbol \Phi(T',\lambda) \Omega(T')dT' \
\
\boldsymbol\Psi(T,\lambda)&=e^{-i\frac{\lambda}{2}\bs J T}-\int_{T}^{\infty} 
e^{-i\frac{\lambda}{2}\bs J(T-T')}\boldsymbol \Psi(T',\lambda) \Omega(T') dT'
\end{align}
\end{subequations}
with the corresponding boundary conditions
\begin{align}
\boldsymbol\Phi(T,\lambda)&\rightarrow e^{-i\frac{\lambda}{2}\bs J T} \quad 
\text{as} \quad T \rightarrow -\infty \\
\boldsymbol\Psi(T,\lambda)&\rightarrow e^{-i\frac{\lambda}{2}\bs J T} \quad 
\text{as} \quad T \rightarrow +\infty.
\end{align}
As these two sets of solutions determine complete sets of eigenfunctions, we 
can relate the two via a matrix $\boldsymbol S(\lambda)$ known as the 
scattering data,
\begin{align}
\boldsymbol\Psi=\boldsymbol\Phi \boldsymbol S(\lambda)\quad \text{where} \quad 
\boldsymbol S(\lambda)=\left( 
\begin{array}{cc}
\bar{\boldsymbol a} &\boldsymbol b\\
\bar{\boldsymbol b}^\dagger & a
\end{array}\right)
\end{align}
with $\bar{\boldsymbol a}$ a $2\times 2$ matrix and $\boldsymbol b$ and $
\bar{\boldsymbol b}$ two-dimensional column vectors. It can be shown that $a$ 
is analytic in the lower-half $\lambda$-plane, and we will also assume that it 
has a finite number of simple zeros, $\lambda_1, ..., \lambda_n$, in the 
region $\text{Im}(\lambda)<0$. The reflection coefficient is given by
\begin{align}
\boldsymbol r (\lambda)=\boldsymbol b(\lambda)/a(\lambda)
\end{align}
which is well-defined on the real $\lambda$-axis.

From Eqs.~\eqref{eq:phisi}, it is clear that the first two columns of $
\boldsymbol \Phi$, which we denote by $\boldsymbol \phi$, and the last column 
of $\boldsymbol \Psi$, denoted by $\boldsymbol \psi$, are analytic in the 
lower-half $\lambda$-plane. Furthermore, they satisfy the following relation:
\begin{align}
\text{det}(\boldsymbol \phi,\boldsymbol \psi) =a(\lambda)e^{i\frac{\lambda}{2}
T}.
\end{align}
Therefore, when evaluated at one of the simple zeros of $a$ they become 
linearly dependent. Thus, we can write
\begin{align}
\boldsymbol \psi(\lambda_j)= \boldsymbol \phi (\lambda_j) \boldsymbol 
\eta^{(j)}.
\end{align}
From this equality, we can define the norming constants, $\boldsymbol
\beta^{(j)}$, which are written in terms of the residues of the quantity $
\boldsymbol \psi /a$,
\begin{align}
\text{Res}\left\{\frac{\boldsymbol \psi}{a},\lambda_j\right\}= \boldsymbol 
\phi(\lambda_j) \boldsymbol \beta^{(j)}, \quad \text{where} \quad \boldsymbol 
\beta^{(j)}=\frac{\boldsymbol \eta^{(j)}}{a'(\lambda_j)}.
\end{align}

In order to solve the CMB, we need to determine the evolution in $Z$ of the 
scattering data (the components of the scattering matrix). This gives a number 
of coupled, nonlinear differential equations which, in general, do not have an 
analytical solution. Fortunately, the inverse scattering procedure can be 
carried out by simply determining the evolution of the norming constants, $
\boldsymbol\beta^{(j)}$, and the reflection coefficient, $\boldsymbol r$, 
along with the location of the zeros of $a$, namely, $\lambda_1, ..., 
\lambda_n$. In spite of this, the general case remains an open problem. Some 
progress in the case of a small $\bs r$ has been presented in 
\cite{chakravarty2014inverse}.

In this manuscript, we restrict ourselves to the reflection-less case, $
\boldsymbol r=0$. This implies that no radiation background will be 
superimposed to the solitonic interaction and that the initial coherences with 
the excited state must be zero. Therefore, the initial density matrix is block 
diagonal and commutes with $\boldsymbol J$. Thus we have
\begin{align}
\label{eq:rho0}
\bs{\rho_0}(Z,\Delta)=
\left(\begin{array}{cc}
\boldsymbol{\rho_g}^{(0)}(Z,\Delta) & 0\\
0&\rho_{33}^{(0)}(Z,\Delta)
\end{array}
\right)
\end{align}
where $\boldsymbol{\rho_g}$ denotes the $2\times2$ density matrix of the 
ground state.

From Eq.~\eqref{eq:scat2}, the evolution of the norming constants can be 
derived and is given by
\begin{align}
\label{eq:beta}
\partial_Z \boldsymbol \beta^{(j)}=\frac{i\mu}{2} \left\langle 
\frac{\boldsymbol\rho_g^{(0)}-\rho_{33}^{(0)}\boldsymbol I_2}{\lambda_j-
\Delta}\right\rangle \boldsymbol \beta^{(j)}.
\end{align}
Solving this equation and using the zeros of $a$ we can compute the fields by 
the following formula
\begin{align}
\label{eq:omegas}
\left(
\begin{array}{c}
\Omega_s\\
\Omega_c
\end{array}
\right)=-4\sum^n_{j,l=1} (\bs{K}^{-1})_{jl} \alpha^*_l \bs \beta^{(j)} e^{i
\lambda_jT}
\end{align}
where 
\begin{align}
\label{eq:K}
K_{ij}=2\frac{\alpha_i^* \alpha_j+ \bs \beta^{(j)T} \bs \beta^{(i)*}
e^{i(\lambda_j-\lambda_i^*)T}}{\lambda_i-\lambda_j^*}
\end{align}
and
\begin{align}
\label{eq:alpha}
\alpha_i=\frac{\prod_{j=1}^n(\lambda_j^*-\lambda_i)}{2\prod_{j\neq i}
(\lambda_j-\lambda_i)}.
\end{align}
Likewise, we can deduce an expression for the density matrix
\begin{align}
\boldsymbol \rho = \boldsymbol \Phi \boldsymbol \rho_0 \boldsymbol \Phi^{-1}.
\end{align}
From  this general expression, we can deduce a simpler formula for the density 
matrix for large $T$, i.e.,~after the pulses have passed, which is given by
\begin{align}
\label{eq:rhof}
\bs \rho^{f}(Z,\Delta)=\left(
\begin{array}{cc}
\left(\bar{ \bs a}^\dagger \bs{\rho_g}^{(0)} \bar{ \bs a}\right)(\lambda=
\Delta) & 0\\
0& \rho_{33}^{(0)}
\end{array}
\right)
\end{align}
with the matrix $\bar{\bs a}$ given by
\begin{align}
\label{eq:ascat}
\bar{\bs a}(\lambda)=\prod_{j=1}^n \bs l_j(\lambda),\quad \bs l_j(\lambda)=\bs 
I_2+\frac{\lambda_j-\lambda_j^*}{\lambda-\lambda_j}\frac{\bs v^{(j)} \bs 
v^{(j)\dagger}}{\|\bs v^{(j)}\|^2} 
\end{align}
where the $\bs v^{(j)}$ vectors are defined recursively as
\begin{align}
\label{eq:vi}
\bs v^{(1)}&=\bs \beta^{(1)}, \quad
\bs v^{(i)}= \left( \prod_{j=1}^{i-1} \bs l_j(\lambda_i) \right)^{-1}\bs 
\beta^{(i)}, \quad i=2,...,3.
\end{align}

In summary, to obtain a solution of the CMB equations (in the reflection-less 
case), first we need to give an initial density matrix of the form given in 
Eq.~\eqref{eq:rho0} and the spectral parameters $\lambda_j$ (simple zeros of 
$a$, the number of spectral parameters determines the order of the solution). 
Then, we compute the evolution of the norming constants by solving Eq.~
\eqref{eq:beta}. Finally, we use Eqs.~(\ref{eq:omegas}-\ref{eq:alpha}) to 
obtain the evolution of the fields and Eqs.~(\ref{eq:rhof}-\ref{eq:vi}) for 
the final state of the atomic medium.

\section{Analytical solution}

\subsection{Pulse storage}
Now that we have all the pieces to compute the solutions, we consider the 
particularly simple case of all the atoms prepared initially in the ground 
state $\ket 1$ which is the most common situation that we encounter when 
dealing with EIT and pulse storage in a $\Lambda$ configuration. Thus, we have
\begin{align}
\bs{\rho_g}^{(0)}=\left( \begin{array}{cc}
1&0\\
0&0
\end{array} \right), \quad \rho_{33}^{(0)}=0.
\end{align}

We will start by considering the one-soliton ($n=1$) solution and write $
\lambda_1=\xi_1-i/\tau_1$ with $\xi_1,\tau_1\in\Re$ and $\tau_1>0$. First, we 
solve Eq.~\eqref{eq:beta} noting that the initial density matrix is 
independent of detuning and the matrix $\boldsymbol\rho_g^{(0)}-\rho_{33}
^{(0)}\boldsymbol I_2$ is diagonal (the two equations for each component of $
\bs \beta$ are decoupled). Thus, it is easy to see that the solution can be 
written as
\begin{equation}
\label{eq:beta1}
\bs \beta^{(1)T}=\left(
c_1e^{-(\kappa_1+i\delta_1)Z},
c_2\right)
\end{equation}
($\bs \beta^{(1)T}$ denotes the transpose of $\bs \beta^{(1)}$) where we 
defined
\begin{subequations}
\label{eq:kappadel}
\begin{align}
\kappa_1=&\frac{\mu}{2 \tau_1}\int \frac{F(\Delta)d\Delta}{(\Delta-\xi_1)^2+
(1/\tau_1)^2},
\end{align}
and
\begin{align}
\delta_1=&\frac{\mu}{2}\int \frac{(\Delta-\xi_1)F(\Delta)d\Delta}{(\Delta-
\xi_1)^2+(1/\tau_1)^2},
\end{align}
\end{subequations}
with $\kappa_1$ being the absorption depth as defined for a weak-field 
excitation.
Now, we use Eq.~\eqref{eq:omegas} to compute the initial pulses. Since $n=1$ 
we have that $\bs K $ is just a scalar and $\alpha=(\lambda_1^*-\lambda_1)/
2=i/\tau_1$. Thus we obtain,
\begin{subequations}
\label{eq:pulses1}
\begin{align}
\Omega_s=&\frac{4c_1}{\tau_1}e^{i(\xi_1 T-\delta_1 Z)} \left[ 2|c_1|\cosh(T/
\tau_1-\kappa_1 Z +\sigma_1)\right.\nonumber\\
&\left.+|c_2| \exp(T/\tau_1+\kappa_1 Z +\sigma_2)\right]^{-1} \\
\Omega_c=&\frac{4c_2}{\tau_1}e^{i\xi_1 T} \left[ 2|c_2|\cosh(T/\tau_1+
\sigma_2)\right.\nonumber\\
&\left.+ |c_1|\exp(T/\tau_1-2\kappa_1 Z +\sigma_1)\right]^{-1} 
\end{align}
\end{subequations}
where we defined
\begin{align}
\sigma_i=\ln (|c_i|\tau_1), \quad i=1,2.
\end{align}
Therefore we see that $\delta_1$ appears as an extra contribution to the 
refractive index for the signal field. 

From these expressions we can calculate the pulse area for each pulse, which 
is defined as 
\begin{equation}
\theta(Z)=\int_{-\infty}^{\infty}|\Omega(Z,T)|dT,
\end{equation}
and find that the two-pulse area \citep{clader2007two} is given by
\begin{equation}
\Theta(Z)=\sqrt{\theta_s(Z)^2+\theta_c(Z)^2}=2\pi.
\end{equation}
This was to be expected as these are essentially the same solutions presented 
in \cite{clader2007two}.
At the boundary, $Z=0$, if $|c^{(1)}|\gg |c^{(2)}|$ these are well 
approximated by 
\begin{subequations}
\label{eq:pulstor}
\begin{align}
\Omega_s=&\frac{\theta_s(Z=0)}{\pi\tau_1}e^{i\xi_1 T} \sech(T/\tau_1+
\sigma_1),\\
\Omega_c=&\frac{\theta_c(Z=0)}{\pi\tau_1}e^{i\xi_1 T} \sech(T/\tau_1+
\sigma_1).
\end{align}
\end{subequations}

\begin{figure}
\centering
\includegraphics[width=1.\linewidth]{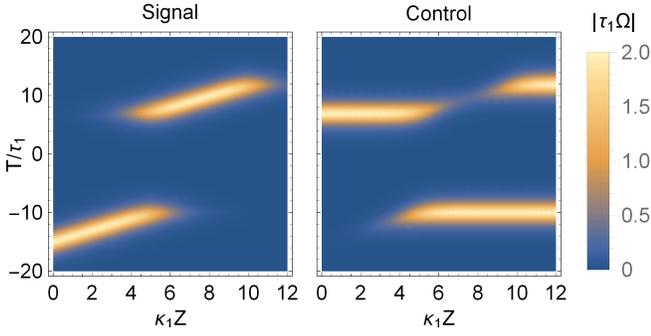}
\caption{\label{fig:pulses} (Color online) Pulse propagation as dictated by 
the second-order solution when $|c^{(1)}|\gg |c^{(2)}|$ and $\sigma_1\gg\zeta
$. The plot on the left (right) shows the evolution of the signal (control) 
pulse. For $T/\tau_1<0$ we have the encoding of the signal pulse onto the 
ground state coherence and for $T/\tau_1>0$ we have the retrieval followed by 
the re-encoding in the displaced location. }
\end{figure}

This solution describes the propagation of two matched pulses with 
complementing areas such that the two-pulse area is always equal to $2\pi$. As 
the signal pulse propagates at a reduced group velocity, it excites some of 
the population to the excited state where it can be stolen by the control 
pulse. This leads to an amplification of the control pulse while the signal 
pulse is absorbed. After the interaction between the two fields, the control 
pulse becomes a full $2\pi$ pulse propagating at the speed of light (as it is 
decoupled from the medium) together with the full absorption of the signal 
field. This is depicted in Fig.~\ref{fig:pulses} for $-20<t/\tau_1<0$.

\begin{figure}
\includegraphics[width=1.\linewidth]{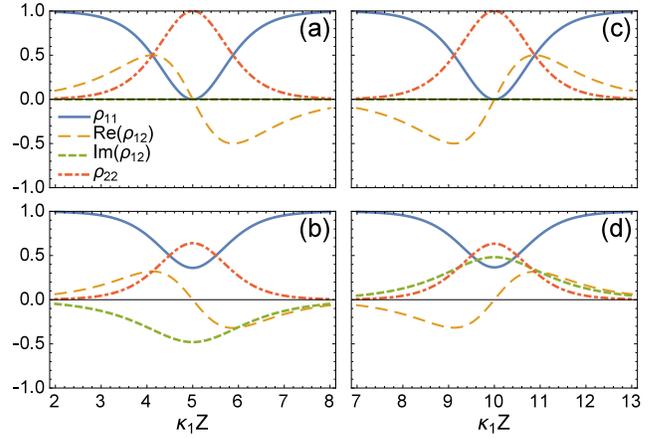}
\caption{\label{fig:den} (Color online) Imprint of the signal pulse in the 
ground state density matrix as it has been encoded before [(a) and (b)] and 
after the displacement [(c) and (d)]. $\tau_1\Delta=0$ for (a) and (c) and $
\tau_1\Delta=0.75$ for (b) and (d).}
\end{figure}

Let us now take a look at the density matrix after the soliton interaction has 
taken place. From Eq.~\eqref{eq:rhof} we have that
\begin{align}
\bs{\rho_g}^{f}(Z,\Delta)=\bar{ \bs a}^\dagger \bs{\rho_g}^{(0)} \bar{ \bs a}=
\left( \begin{array}{cc}
|\bar a_{11}|^2 & \bar a_{11}^*\bar a_{12}\\
\bar a_{11}\bar a_{12}^*&|\bar a_{12}|^2 
\end{array}\right).
\end{align}
and using Eqs.~\eqref{eq:ascat} and \eqref{eq:vi} we obtain
\begin{subequations}
\begin{align}
\bar a_{11}(\lambda)&=1-\frac{i/\tau_1}{\lambda-\xi_1+i/\tau_1}(1-
\tanh(\kappa_1 Z -\sigma_{12})), \\
\bar a_{12}(\lambda)&=-\frac{i/\tau_1}{\lambda-\xi_1+i/\tau_1}C_{12}e^{-i
\delta_1 Z}\sech(\kappa_1 Z -\sigma_{12}),
\end{align}
\end{subequations}
where we defined 
\begin{align}
\sigma_{12}=\ln |c_1/c_2|, \quad \text{and} \quad C_{12}=\frac{c_1c^{*}_2}{|
c_1c_2|}.
\end{align}
Finally, we can write
\begin{subequations}
\label{eq:rhos}
\begin{align}
\rho_{11}=&1+\frac{1}{\tau_1^2(\Delta-\xi_1)^2+1}\left[\tanh^2(\kappa_1 Z -
\sigma_{12})-1\right] ,\\
\rho_{22}=& \frac{1}{\tau_1^2(\Delta-\xi_1)^2+1}\sech^2(\kappa_1 Z -
\sigma_{12}),\\
\rho_{12}=&-\frac{C_{12}e^{-i\delta_1 Z}}{\tau_1^2(\Delta-\xi_1)^2+1}\left[i
\tau_1(\Delta-\xi_1)+\tanh(\kappa_1 Z -\sigma_{12})\right] \nonumber\\
&\times \sech(\kappa_1 Z -\sigma_{12}).
\end{align}
\end{subequations}
From these equations it is clear that while the switching between the signal 
and control pulses is taking place,  the signal pulse is being transferred 
into a spin wave (imprint) located at $\kappa_1 Z=\sigma_{12}$. This location, 
denoted $x_1$, can be written in terms of the pulse area to give the simple 
formula
\begin{equation}
\label{eq:x1}
\kappa_1 x_1=\ln\left(\frac{\theta_s(Z=0)}{\theta_c(Z=0)}\right).
\end{equation}
Hence, the ratio of the pulse areas plays a key role during the storage of the 
signal pulse as it determines the location of the imprint. Figures 
\ref{fig:den}(a) and (b) show the shape of the imprint for two different 
detuning values. 
The storage procedure is in line with the proposal made in 
\cite{grigoryan2009short}, where an intense pulse is mapped into a spin wave 
by means of a low-intensity control pulse in order to reduce the storage 
length as compared to the usual EIT techniques. Yet, no mention of a retrieval 
procedure was made, and so we address this point in the next section.

\subsection{Pulse retrieval and spin wave manipulation}

Having analyzed in detail how an intense pulse can be mapped into a spin wave, 
we need to address the matter of retrieving the information stored. Looking at 
previous work on the subject, we see that in order to recover the signal pulse 
we need to apply a control field. Going back to the first-order solution we 
notice that if we set the integration constant $c_1$ in \eqref{eq:beta1} equal 
to zero, the norming constant is given by
\begin{align}
\label{eq:beta2}
\bs \beta^{(2)T}=(0,d).
\end{align}
This leads to a $2\pi$-control field propagating at the speed of light in 
vacuum
\begin{align}
\Omega_c=&\frac{2}{\tau_2}\frac{d}{|d|}e^{i\xi_2 T} \sech(T/\tau_2+\zeta),
\end{align}
where the spectral parameter was taken to be  $\lambda_2=\xi_2-i/\tau_2$ and 
we defined
\begin{equation}
\zeta=\ln (|d|\tau_2).
\end{equation}

To study how this control pulse interacts with the spin wave encoded in the 
medium, we compute the second-order ($n=2$) solution obtained from the norming 
constants as defined in Eqs.~\eqref{eq:beta1} and \eqref{eq:beta2}, and the 
two spectral parameters $\lambda_1$ and $\lambda_2$. Using Eqs.~
\eqref{eq:omegas} through \eqref{eq:alpha} we obtain an exact solution which 
is too cumbersome to be of any use. Hence, we look at the limiting case of $|
c^{(1)}|\gg |c^{(2)}|$ and $\sigma_1\gg\zeta$ for which the pulses at the 
boundary $Z=0$ can be well approximated by
\begin{subequations}
\label{eq:pulses2}
\begin{align}
\Omega_s=&\frac{2}{\tau_1}\frac{c_1}{\sqrt{|c_1|^2+|c_2|^2}}e^{i\xi_1 T} 
\sech(T/\tau_1+\sigma_1)\nonumber\\
&+\frac{2}{\tau_2}\frac{c_2}{\sqrt{|c_1|^2+|c_2|^2}}e^{i\xi_2 T} \sech(T/
\tau_2+\zeta-\chi),\\
\Omega_c=&\frac{2}{\tau_1}\frac{c_2}{\sqrt{|c_1|^2+|c_2|^2}}e^{i\xi_1 T} 
\sech(T/\tau_1+\sigma_1)\nonumber\\
&+\frac{2}{\tau_2}\frac{c_1}{\sqrt{|c_1|^2+|c_2|^2}}e^{i\xi_2 T} \sech(T/
\tau_2+\zeta-\chi),
\end{align}
\end{subequations}
where we defined the phase-lag parameter
\begin{equation}
\label{eq:xi}
\chi=\frac{1}{2}\ln\left( \frac{(\tau_1+\tau_2)^2+(\xi_1-
\xi_2)^2\tau_1^2\tau_2^2}{(\tau_1-\tau_2)^2+(\xi_1-\xi_2)^2\tau_1^2\tau_2^2} 
\right).
\end{equation}

This expression describes a well-defined sequence of pulses. First, we have 
the matched pulses of duration $\tau_1$ that encode the strong signal pulse 
into the medium. These have the same form as the pulses in Eq.~
\eqref{eq:pulstor} and thus lead to the storage of the signal pulse at a 
location determined by Eq.~\eqref{eq:x1}. Then we have another set of matched 
pulses but of duration $\tau_2$ and inverted areas, that is, we have a strong 
control pulse and a weak signal pulse having the same areas as the initial 
storing set.
This solution is shown in Fig.~\ref{fig:pulses}, where we can clearly see that 
initially we have the same type of solution described in the previous section 
followed by the interaction of the second set of matched pulses with the spin 
wave left by the first. As the pulses approach the location of the imprint, 
the signal pulse is amplified at the expense of the control pulse up to the 
point where it becomes a full $2\pi$ pulse (here we can say that the signal 
pulse is retrieved). After propagating a given distance, we have another 
reversal of intensities: the signal pulse is absorbed leading to the 
amplification of the control pulse which, after becoming a full $2\pi$ pulse, 
becomes decoupled from the medium. 

There are two main points to mention as far as the usual concept of pulse 
storage and retrieval. First, the retrieved pulse does not have the same 
properties as the original signal pulse if the spectral parameters differ. The 
retrieval would, of course, be achieved by cutting off the medium at the 
location where the signal pulse has maximum intensity. Second, the retrieval 
control pulse is not just a control field but is accompanied by the same 
amount of signal field as there was of the control field for the storage. This 
contrasts with the usual distinction of one frequency controlling the other. 
We will study how this compares with just using a $2\pi$ control field for the 
retrieval in Sec.~\ref{sec:num}.

Now let us take a look at the density matrix after the interaction of the 
second matched pulse set with the spin wave left behind by the first. For 
this, we again use Eqs.~\eqref{eq:ascat} and \eqref{eq:vi} to obtain the 
following expressions for scattering coefficients:
\begin{subequations}
\begin{align}
\bar a_{11}(\lambda)&=1-\frac{i/\tau_1}{\lambda-\xi_1+i/\tau_1}(1-
\tanh(\kappa_1 Z -\sigma_{12}-\chi)), \\
\bar a_{12}(\lambda)&=-\frac{i/\tau_1}{\lambda-\xi_1+i/\tau_1}C_{12}e^{-i
\delta Z}e^{i\phi}\sech(\kappa_1 Z -\sigma_{12}-\chi),
\end{align}
\end{subequations}
where the extra phase $\phi$ depends on the two eigenvalues $\lambda_1$ and $
\lambda_2$. When these are purely imaginary, $\phi=\pi$. As a result, the 
density matrix takes the form
\begin{subequations}
\begin{align}
\rho_{11}=&1+\frac{1}{\tau_1^2(\Delta-\xi)^2+1}\left[\tanh^2(\kappa_1 Z -
\sigma_{12}-\chi)-1\right] ,\\
\rho_{22}=& \frac{1}{\tau_1^2(\Delta-\xi)^2+1}\sech^2(\kappa_1 Z -\sigma_{12}-
\chi),\\
\rho_{12}=&\frac{-C_{12}e^{-i\delta Z}e^{i\phi}}{\tau_1^2(\Delta-\xi)^2+1}
\left[i\tau_1(\Delta-\xi)\right.\left.+\tanh(\kappa_1 Z -\sigma_{12}-\chi)
\right]\nonumber\\
&  \times \sech(\kappa_1 Z -\sigma_{12}-\chi).
\end{align}
\end{subequations}
These have the same form as those given in Eqs.~\eqref{eq:rhos}, and so the 
result of the interaction with the second set is clear: the spin wave has been 
displaced by an amount determined by the phase-lag parameter $\chi$. 
Additionally, the coherence suffered a $\phi$ phase shift. Therefore, if the 
medium is not cut off before the retrieved pulse is absorbed, the imprint is 
displaced further into the medium with new location 
\begin{equation}
\label{eq:x2}
\kappa_1 x_2=\sigma_{12}+\chi.
\end{equation}

\subsection{Effects of Doppler broadening and detuning}

Let us take a moment to study the effect of inhomogeneous broadening and 
detuning in the solutions just described. Tracing back our steps, we readily 
notice that the only place where a Doppler average is taken is in Eqs.~
\eqref{eq:kappadel}. From the expressions for the pulses and the spin-wave it 
is clear that the absorption coefficient sets the spatial dimension, that is, 
the smaller $\kappa_1$ is, the longer the atomic sample will need to be in 
order to fit the pulses and the spin-wave. On the other hand, the only effect 
of the parameter $\delta_1$ is to produce a position-dependent phase to the 
signal pulse which is in turn transferred to the spin-wave.

So far we have kept the spectral parameters complex, but taking a look at the 
expression for the pulses [Eqs.~\eqref{eq:pulses1} and \eqref{eq:pulses2}] we 
notice that the imaginary part can be identified as the pulse duration while 
the real part appears as a self-detuning term. This term also appears in the 
expression for $\kappa_1$ and $\delta_1$ by shifting the detuning everywhere 
except in the Doppler distribution. Furthermore, looking back at the 
definition of the Rabi frequencies and the slow-varying envelopes for the 
fields, we see that this self-detuning should be included as a real detuning. 
Thus we set it to zero and rewrite the expressions for $\kappa_1$ and $
\delta_1$ in term of the absorption coefficient in the absence of Doppler 
broadening, $\kappa_0=\mu \tau_1/2$,
\begin{subequations}
\label{eq:norm}
\begin{align}
\kappa_1=&\kappa_0\int \frac{\bar F(\nu)d\nu}{\nu^2+1}, \\
\delta_1=&\kappa_0\int \frac{\nu \bar F(\nu)d\nu}{\nu^2+1},
\end{align}
\end{subequations}
where
\begin{align}
\bar F(\nu)= \frac{T_2^\star}{\tau\sqrt{2 \pi}}e^{-(\nu-\tau \bar
\Delta)^2T_2^{\star 2}/2 \tau^2}.
\end{align}

\begin{figure}
\includegraphics[width=1.\linewidth]{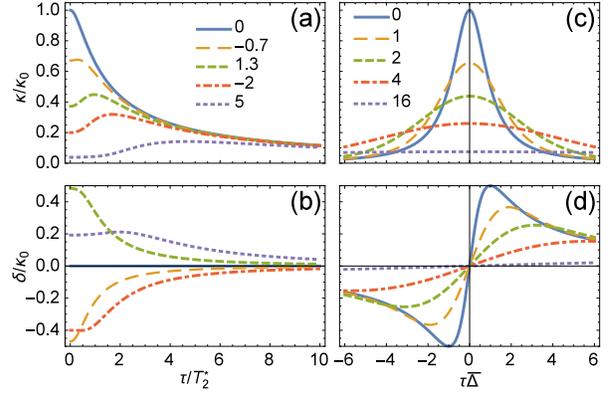}
\caption{\label{fig:dopp} (Color online) Absorption coefficient $\kappa$ and 
extra contribution to the index of refraction $\delta$ as a function of the 
Doppler width and mean detuning. (a) shows $\kappa$ and (b) $\delta$ as a 
function of $\tau/T_2^*$ for $\tau \bar \Delta=0,-0.6,1.2,-2,5$. (c) shows $
\kappa$ and (d) $\delta$ as a function of $\tau \bar \Delta$ for $\tau/
T_2^*=0,1,2,4,16$. }
\end{figure}

We plot these as a function of the Doppler distribution width and mean (see 
Fig.~\ref{fig:dopp}). We readily notice that the maximum value for the 
absorption coefficient is attained in the absence of Doppler broadening and 
zero detuning. When the detuning is non-zero, the maximum for $\kappa$ shifts 
to a wider Doppler distribution. As the absorption coefficient sets the 
spatial length, this can be used to reduce the size of the atomic sample if 
detuning is unavoidable. As for the extra contribution to the index of 
refraction, there must be some detuning in order for it to have some effect 
due to the anti-symmetry of $\delta_1$ with respect to $\bar \Delta$. 
Moreover, both parameters tend to zero in the limits of large width and 
detuning, so in general it is preferable to keep them as low as possible to 
reduce the size of the atomic sample.

The absorption coefficient also determines the group velocity of the signal 
pulse, which is given by
\begin{equation}
\frac{v_g}{c}=\frac{1}{1+\kappa_1 c \tau}.
\end{equation}
Hence, as $\kappa$ becomes smaller the signal pulse travels faster through the 
medium and so needs to interact for a larger distance with the medium to be 
able to produce the same effect as a slower pulse. This is in clear accordance 
with the previous discussion about how $\kappa$ sets the spatial dimension. 
The change in spatial dimension is also present in the expressions for the 
location of the initial imprint and the displacement which can be written as
\begin{subequations}
\begin{align}
\kappa_0 x_1=\frac{\kappa_0}{\kappa_1}\ln\left(\frac{\theta_s(Z=0)}
{\theta_c(Z=0)}\right), \\
\kappa_0 \Delta x=\kappa_0 (x_2- x_1)=\frac{\kappa_0}{\kappa_1}\ln\left( 
\frac{\tau_1+\tau_2}{\tau_1-\tau_2} \right).
\end{align}
\end{subequations}

\section{Non-ideal pulse storage and retrieval}
\label{sec:num}

After discussing the analytic solution in detail we need to compare it to more 
realistic conditions. In order to accomplish this we recur to numerical 
calculations. Note that the initial assumption of zero decay was justified 
since the lifetime of the higher level is around two orders of magnitude 
longer that the duration of the pulses for the regime in which we are working. 
As an example, for Rb we have $\tau \Gamma\approx 0.01$ for pulse duration $
\tau \approx 0.26 \text{ns}$ \citep{steck2001rubidium}. We will consider it, 
as well as the finiteness of the pulses, as it may have a noticeable effect. 
We will assume that the decay rate into each ground state is the same for a 
given decay rate $\Gamma$ for the excited state. Additionally, we set the Rabi 
frequencies equal to zero when $|\tau_1\Omega|<10^{-5}$.

\begin{figure}
\includegraphics[width=1.\linewidth]{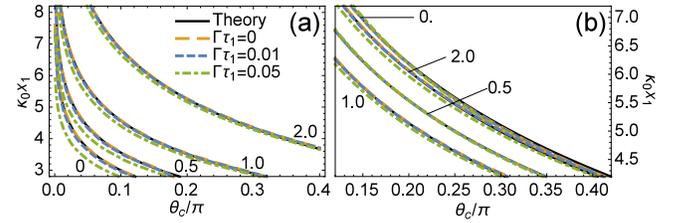}
\caption{\label{fig:x1} (Color online) Location of the stored spin wave as a 
function of the control pulse area for (a) $\tau \bar \Delta=0$ and $\tau/
T_2^*= 0,0.5,1,2$ and (b) $\tau \bar \Delta=1.3$ and $\tau/T_2^*=0, 0.5,1,2$.}
\end{figure}

First, we consider the storage of the signal pulse and thus the location of 
the stored spin-wave. The results for different values of the parameters are 
shown in Fig.~\ref{fig:x1}. We notice that it is difficult to distinguish 
between the different curves, but that is due to the agreement between the 
analytical solution and the numerical results. In the absence of detuning 
[Fig.~\ref{fig:x1}(a)] we note that the effect of spontaneous emission is to 
lower the curve, that is, the signal pulse is stored before the predicted 
location. This is the same effect that was reported in 
\cite{gutierrez2015manipulation} in the absence of Doppler broadening.

If we now take a closer look at Fig.~\ref{fig:x1}(b) we can spot some 
differences. The most noticeable effect is that the curves have been reversed. 
In this case ($\tau \bar \Delta=1.3$) the spin wave for a Doppler width of $
\tau/T_2^*= 0.5$ is located after the one made for $\tau/T_2^*= 1$. This is 
just a consequence of the displaced maximum for $\kappa_1$ as a function of 
the width [see Fig.~\ref{fig:dopp}(a)]. In addition there is a somewhat hidden 
feature. For a Doppler width of $\tau/T_2^*= 0.5$ the effect of spontaneous 
emission is suppressed: the four curves overlap almost perfectly. While for $
\tau/T_2^*= 0$ the effect is reversed. As for the other two width values, the 
curves with spontaneous emission are lowered, but for  $\tau/T_2^*=1$ the 
effect is less noticeable. 

\begin{figure}
\includegraphics[width=1.\linewidth]{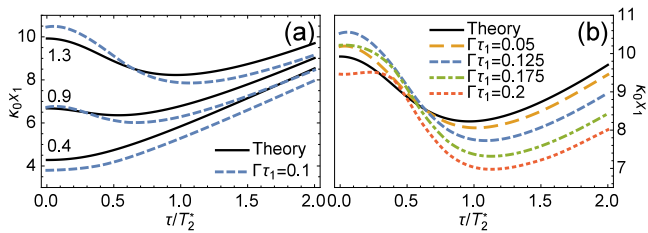}
\caption{\label{fig:x1t2} (Color online) Location of the stored spin wave as a 
function of the Doppler width for a control pulse area of $\theta_c=0.05\pi$. 
In (a) we fix the decay rate and consider different values for the detuning, $
\tau \bar \Delta=0.4,0.9,1.3$ and in (b) we fix the detuning at $\tau \bar 
\Delta=1.3$ and vary the decay rate. }
\end{figure}

To investigate this strange phenomenon we plot the location of the imprint as 
a function of $T_2^*$ for a higher value of spontaneous emission $\tau 
\Gamma=0.1$ [see Fig.~\ref{fig:x1t2}(a)]. The effect is clear: for 
sufficiently detuned fields the effect of spontaneous emission is reversed for 
small values of the Doppler width. Now, with increasing rate of spontaneous 
emission, the reversal is enhanced until it reaches a maximum [see Fig.~
\ref{fig:x1t2}(b)]. After this, the effect is diminished. Comparing the 
crossing of the numerical results with the curve predicted by the theory we 
notice that this reversal is limited to the region where $\partial \kappa/
\partial (\tau/T^*_2) >0$. This effect is independent of the sign of the 
detuning.

\begin{figure}[t]
\includegraphics[width=1.\linewidth]{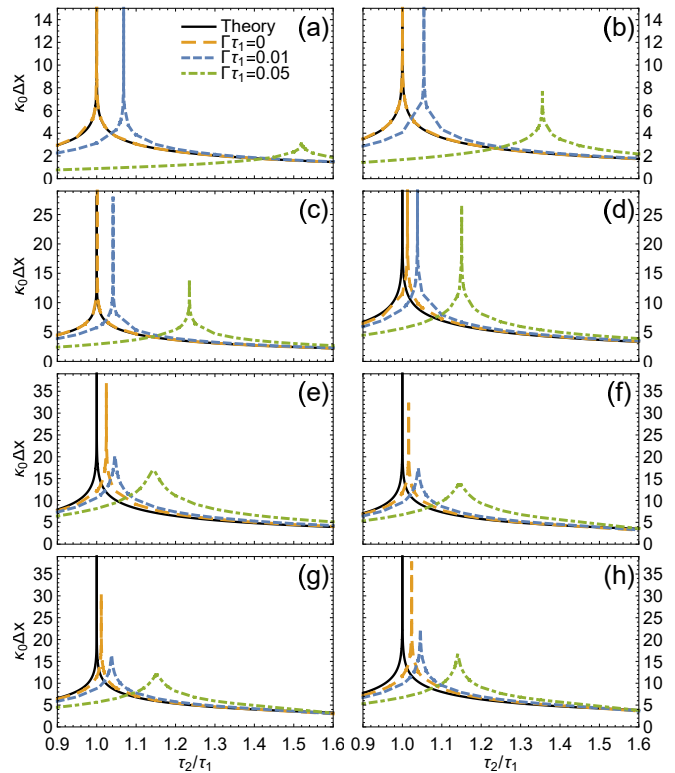}
\caption{\label{fig:dis} (Color online) Displacement of the stored spin wave 
as a function of the second control pulse duration for  $\tau \bar \Delta=0$ 
and (a) $\tau/T_2^*= 0$, (b) $\tau/T_2^*= 0.5$, (c) $\tau/T_2^*= 1$ and (d) $
\tau/T_2^*= 2$. Then we fix $\tau \bar \Delta=1.3$ and (e) $\tau/T_2^*= 0$, 
(f) $\tau/T_2^*= 0.5$, (g) $\tau/T_2^*= 1$ and (h) $\tau/T_2^*= 2$. The 
initial location of the imprint was chosen to be at $\kappa_0 x_1=5$ where the 
storage was made by a $2\pi$-signal pulse and the required control pulse.}
\end{figure}

Now, let us take a look at the displacement of the spin-wave. We show the 
results for different values of Doppler width, detuning and spontaneous decay 
rate along with the theoretical result in Fig.~\ref{fig:dis}. We notice that 
in the absence of spontaneous emission the numerical results are very similar 
to those predicted by the analytical solution. The only difference is that the 
maximum displacement is attained for values slightly bigger than $\tau_2=
\tau_1$. The biggest deviation is when $\tau \bar \Delta=1.3$ and $\tau/
T_2^*=2$ [see  Fig.~\ref{fig:dis}(h)], which is also the case where the 
largest control-pulse area is required for the storage of the $2\pi$-signal 
pulse and thus is the one where the two-pulse area deviates the most from the 
one predicted for the storage stage.

\begin{figure}
\includegraphics[width=1.\linewidth]{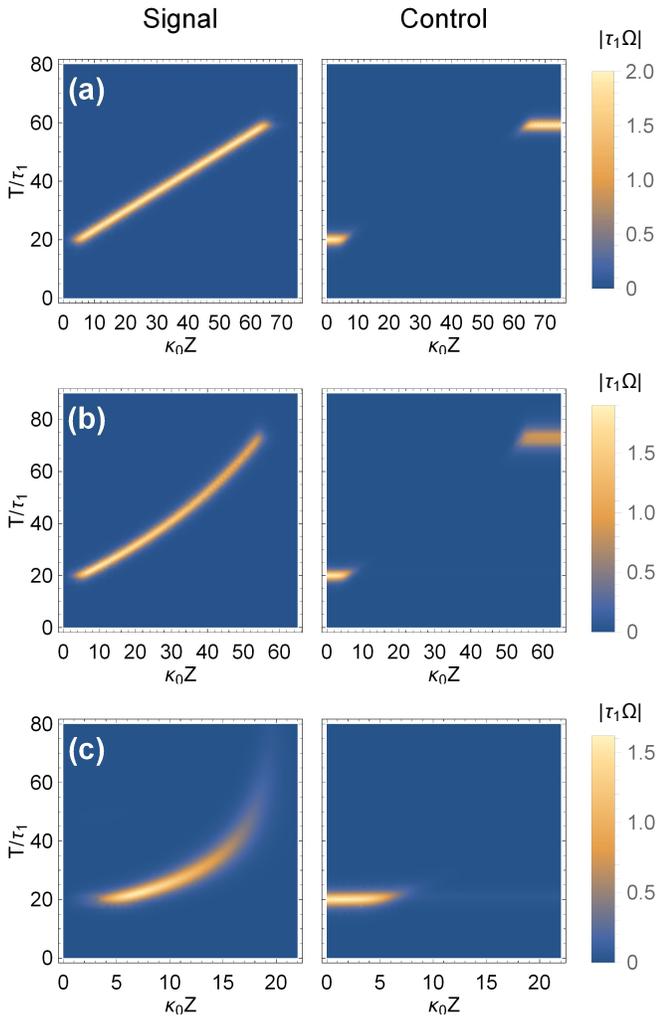}
\caption{\label{fig:retdyn} (Color online) Pulse dynamics around the 
displacement peaks for  $\tau \bar \Delta=0$ and (a) $\tau \Gamma= 0$, (b) $
\tau \Gamma= 0.01$, and (c) $\tau \Gamma= 0.05$.}
\end{figure}

The effect of spontaneous emission is evident: as the decay rate increases, 
the maximum displacement is shifted to larger values of $\tau_2$. This is 
coupled with a lowering in the maximum displacement with increasing rate of 
spontaneous emission: For the cases with no detuning and spontaneous decay $
\tau \Gamma \leq 0.01$, by using double precision numbers, no maximum was 
attained but we can be fairly certain that a maximum should be attained at 
least for the cases with non-zero decay because of the pulse dynamics shown in 
Fig.~\ref{fig:retdyn} as explained in the next paragraph. Here, there is a 
clear difference between the cases with spontaneous emission and the one 
without, unlike in the storage stage. Nonetheless, a considerable displacement 
of the spin-wave can be achieved in each case as long as the right duration 
for the control pulse is chosen (this feature was not reported in 
\citep{gutierrez2015manipulation} as the duration of the control pulses was 
always lower than that of the original signal pulse). 

\begin{figure}
\includegraphics[width=1.\linewidth]{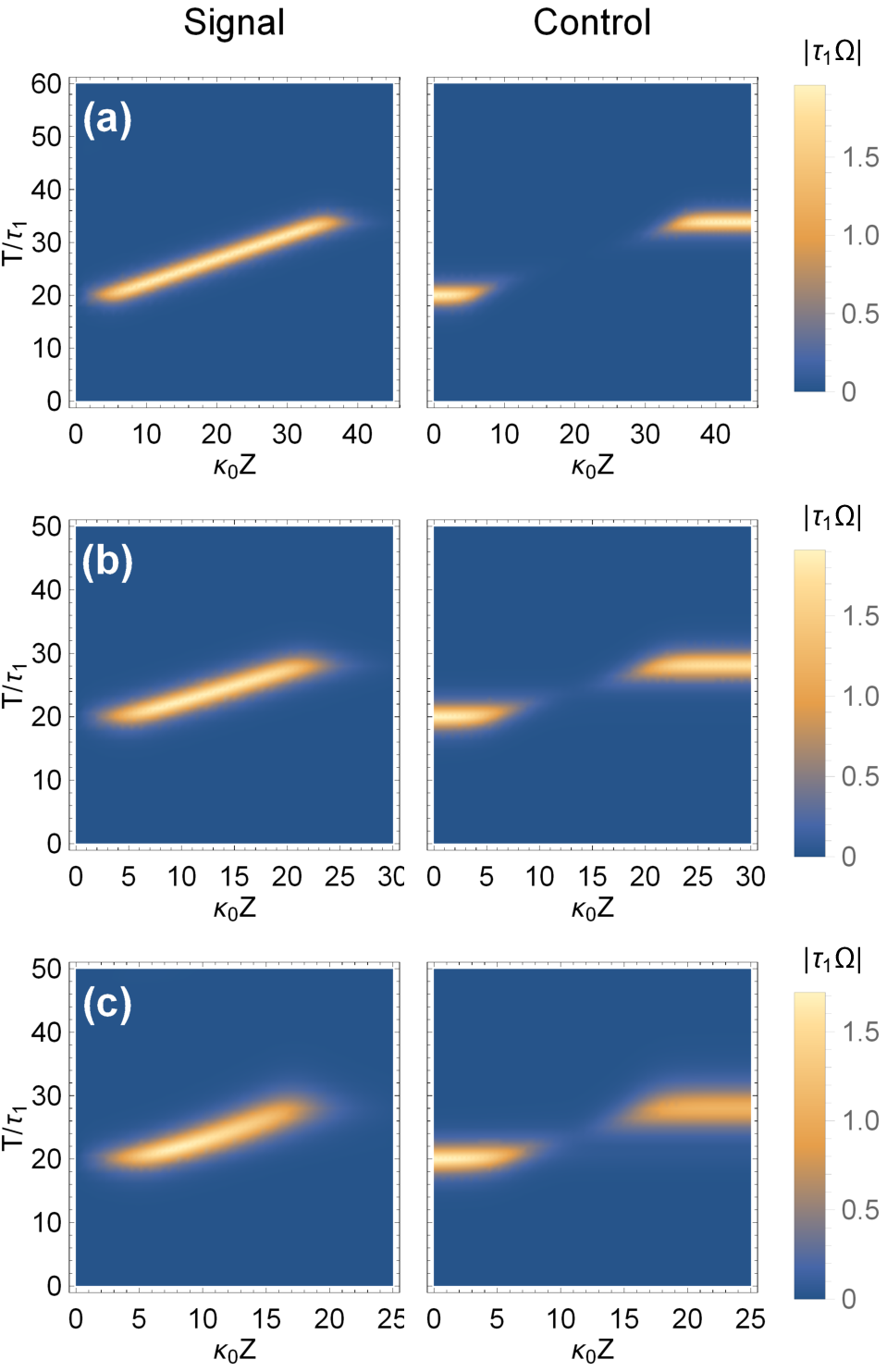}
\caption{\label{fig:retdyndet} (Color online) Pulse dynamics around the 
displacement peaks for  $\tau \bar \Delta=1.3$ and (a) $\tau \Gamma= 0$, (b) $
\tau \Gamma= 0.01$, and (c) $\tau \Gamma= 0.05$.}
\end{figure}

\begin{figure}
\includegraphics[width=1.\linewidth]{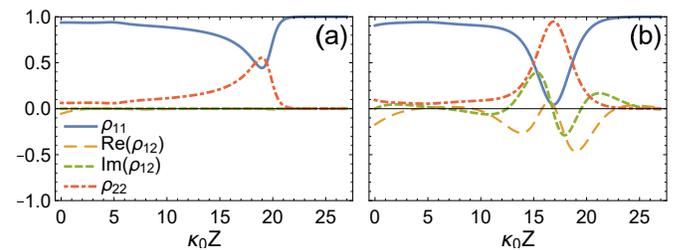}
\caption{\label{fig:retden} (Color online) Density matrix elements for the 
ground states after the maximum displacement for $\tau \Gamma= 0.05$ and (a)  
$\tau \bar \Delta=0$ and (b)  $\tau \bar \Delta=1.3$.}
\end{figure}
Figure \ref{fig:retdyn} shows the pulse dynamics in the retrieval stage at the  
displacement peaks for the cases in resonance with $\tau/T_2^*= 1$. We only 
show one value for the Doppler broadening width as the other cases behave in 
the same way. Let us first take a look at the case in resonance. The first 
thing we notice is that for the case with no decay, the retrieval behaves 
pretty much as described by the analytical solution.  Now, if we add some 
decay channel we notice that the signal pulse starts slowing down at the same 
time as it decays. For the case with highest spontaneous emission the signal 
pulse almost comes to a full stop, as can be understood by the close to 90 
degrees turn. As for the case with $\tau \Gamma= 0.01$, the slowing down and 
decay are evidence of the eventual maximum displacement which was not achieved 
due to limitations in the precision. If we now go off resonance (see Fig.~
\ref{fig:retdyndet}) we notice that the slowing down disappears. This can be a 
result of the reduced population induced in the excited state. Another 
consequence of this is the preservation of the coherence between the ground 
state levels which would allow another retrieval. Figure \ref{fig:retden} 
shows the  density matrix elements for the cases with $\tau \Gamma= 0.05$. We 
readily notice that the coherence is completely lost when in resonance whilst 
it is mostly preserved in the detuned scenario. Therefore, the detuning 
prevents the loss of coherence due to spontaneous emission. It is also worth 
noting that a maximum displacement is always attained even for the case with 
no decay channel.

\begin{figure}
\includegraphics[width=1.\linewidth]{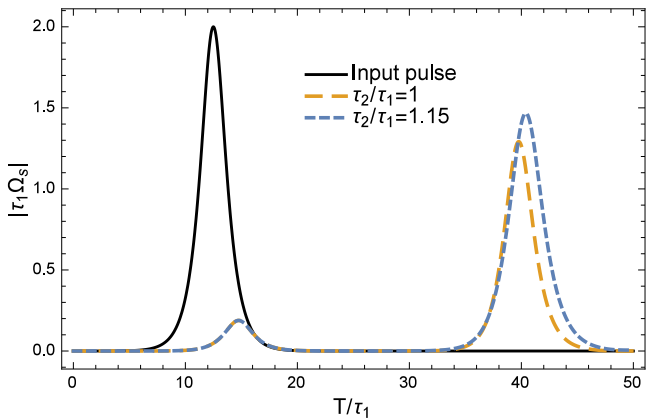}
\caption{\label{fig:retr} (Color online) Input $2\pi$-signal pulse ($\kappa_0 
Z=0$) and output pulses ($\kappa_0 Z=10$) for two values of the duration of 
the control pulse $\tau_2/\tau_1=1,1.15$. The atomic medium is ten $
\kappa_0^{-1} $ long and we chose $\tau \bar \Delta=0$, $\tau/T_2^*=2$ and $
\tau_1\Gamma=0.05$.  }
\end{figure}

Moving on to the retrieval of the signal pulse we find that there is a caveat, 
though. As we already mentioned in the analytical solution, the retrieved 
signal pulse inherits the duration of the control pulse used to retrieve it. 
This remains true in the non-idealized scenario as can be seen in Fig.~
\ref{fig:retr}. Therefore there is a compromise: if we want the highest output 
intensity then we have to allow the duration of the retrieved pulse to be 
different or, conversely, if we want to keep the same parameters for the 
signal pulse we will have to settle for a lower output intensity. In this 
regard, the presence of Doppler broadening and detuning helps by bringing the 
maximum displacement closer to that predicted by the theory.

As a last note, this scheme also helps reduce the storage length compared to 
the usual EIT scheme. This was also noted in \cite{grigoryan2009short}. For 
the parameters considered in this work the EIT-type interaction has a minimal 
effect in both the control and signal fields for the equivalent distances 
considered here. These need to be substantially increased in order to see the 
formation of the adiabaton pair \citep{grobe1994formation}.

\section{Conclusions}

In conclusion, we have used the technique of inverse scattering to obtain 
analytical solutions that describe the storage and retrieval of intense pulses 
in the presence of Doppler broadening and detuning. These results are in 
accordance with previous treatments 
\citep{groves2013jaynes,gutierrez2015manipulation}. Only minor corrections are 
needed and the main effect (according to analytical solutions) is the change 
in length scale which is determined by $\kappa_1$, thus showing that the 
preliminary conjectures made in \citep{gutierrez2015multi} were somewhat 
pessimistic.

Additionally, we have tested the result given by the analytical solution via 
numerical solutions through which we have found a notable agreement for the 
storage step. The addition of homogeneous broadening causes a minimal 
deviation from the theoretical predictions and a strange reversal of its 
effect in the region where $\partial \kappa/\partial (\tau/T^*_2) >0$. As for 
the displacement (or retrieval, if the displacement is greater than the medium 
length), the differences with the theory are appreciable but nonetheless 
follow a similar trend. Most noticeable is the shift of the maximum for the 
displacement of the spin-wave (an effect previously overlooked) which would 
allow the experimental realization of this procedure even for higher values of 
decay rate. It is also worth highlighting the slowing down of the retrieved 
pulse when in resonance (due to the spontaneous decay) and its disappearance, 
along with the protection of coherence as we go off-resonance.

The results presented here and in \citep{gutierrez2015manipulation} show that 
this storage scheme can be implemented experimentally in a variety of 
conditions. In addition, the backward-transfer solution as well as the multi-
pulse storage presented in \citep{gutierrez2015multi} should remain valid with 
the appropriate change of the spatial dimension.

\begin{acknowledgments}
The author gratefully acknowledges J.~H.~Eberly and M.~A.~Alonso for fruitful 
discussions and a careful review of this manuscript.
This work was supported by  NSF grants (PHY-1203931, PHY-1505189) and a 
CONACYT fellowship 
awarded to R. Guti\'{e}rrez-Cuevas.  
\end{acknowledgments}


%

\end{document}